# AI-Driven Drug Repurposing through miRNA-mRNA Relation

Sharanya Manoharan[1 †], Balu Bhasuran[2 †], Oviya Ramalakshmi Iyyappan[3], Mohamed Saleem Abdul Shukkoor[4], Malathi Sellapan[5], Kalpana Raja[6]*

[1]Department of Bioinformatics, Stella Maris College, Chennai, Tamil Nadu, India; sharanya.bioinfo@gmail.com
[2]School of Information, Florida State University, Tallahassee, FL, USA; bbhasuran@fsu.edu
[3]Department of Computer Science and Engineering, Amrita School of Computing, Amrita Vishwa Vidyapeetham, Chennai, Tamil Nadu, India; iroviya@gmail.com
[4]College of Pharmacy, Riyadh Elm University, Riyadh, Kingdom of Saudi Arabia; saleemskma@yahoo.com
[5]Department of Pharmaceutical Analysis, PSG College of Pharmacy, Coimbatore, Tamil nadu, India; malathisanju@gmail.com
[6]Department of Biomedical Informatics and Data Science, School of Medicine, Yale University, New Haven, CT, USA

[†] Authors contributed equally

*Correspondence: kalpana.raja@yale.edu

**Abstract**

miRNA-mRNA relations are closely linked to several biological processes and disease mechanisms. In a recent study, we tested the performance of large language models (LLMs) on extracting miRNA-mRNA relations from PubMed. PubMedBERT achieved the best performance of 0.783 F1-score for miRNA–mRNA Interaction Corpus (MMIC). Here, we first applied the fine-tuned PubMedBERT model to extract miRNA-mRNA relations from PubMed for chronic obstructive pulmonary disease (COPD), Alzheimer's disease (AD), stroke, type 2 diabetes mellitus (T2DM), chronic liver disease, and cancer. Next, we retrieved miRNA-drug relations using KinderMiner, a literature mining tool for relation extraction. Then, we constructed three interaction networks, *(1)* disease centric network, *(2)* drug centric network, and *(3)* miRNA centric network, comprising 3,497 nodes and 16,417 edges organized as a directed graph to capture complex biological relationships. Finally, we validated the drugs using MIMIC IV. Our integrative approach revealed both established and novel candidate drugs for diseases under study through 595 miRNA-drug relations extracted from PubMed. To the best of our knowledge, this is the first study to systematically extract and visualize relationships among four distinct biomedical entities, miRNA, mRNA, disease, and drug, laying the foundation for repurposing drugs for any disease of interest.

## Introduction

Gene-targeted therapy has gained significant attention due to its potential to treat a range of diseases. A comprehensive understanding of disease-associated RNAs, both coding (mRNA) and non-coding RNAs (lncRNA, miRNA, siRNA, circRNA, piRNA, etc.), in normal and diseased states provide critical insights into the molecular complexity of diseases. Among these, miRNA has emerged as a potential biomarker for disease diagnosis and therapeutic development. Studies show that miRNAs regulate approximately 60% of protein-coding genes [1] and influence key cellular processes, including cell proliferation, apoptosis, and necrosis [2,3,4]. miRNA-mRNA interaction exhibits a negative correlation, where miRNA binds to specific seed regions of target mRNAs. The interaction leads to mRNA degradation or translational repression, thereby modulating protein levels and cellular functions [5,6].

Dysregulation of miRNA-mRNA interactions disrupts normal cell function, which leads to diseases such as cancer, cardiovascular disorders, neurological conditions, and immune-related abnormalities. In cancer, overexpression of oncogenic miRNAs (oncomiRs) or downregulation of tumor-suppressive miRNAs can alter the genes controlling proliferation, apoptosis, and metastasis, thus promoting tumorigenesis [7]. In neurological disorders like Alzheimer's disease (AD), abnormal miRNA-mRNA interactions impair synaptic plasticity, increase neuroinflammation, and comprehensive neuronal survival [8]. In pulmonary diseases like chronic obstructive pulmonary disease (COPD), miRNA-mRNA interactions contribute to disease progression, and therapeutic response.

Knowledge on miRNA-drug association is increasingly important for drug repurposing strategies. Drugs can modulate specific miRNA to restore normal gene expression [9,10]. For instance, trichostatin A altered the expression of 32 miRNAs in breast cancer cell lines [11]. The specific secondary structures and conserved sequences of miRNAs make them attractive drug targets, capable of modulating multiple genes simultaneously. MRX34, the first cancer-targeted miRNA drug, entered Phase I clinical trials for advanced hepatocellular carcinoma in 2013 [12].

In a recent study, we fine-tuned large language models (LLMs) using miRNA-mRNA Interaction Corpus (MMIC), an annotated corpus that includes 1,000 sentences with miRNA-mRNA relations mentioned in PubMed abstracts. Among the machine learning, deep learning and LLM-based approaches tested for extracting miRNA-mRNA interactions from PubMed sentences, PubMedBERT achieved the best performance of 0.786 F-score [13]. In the current study, we used (i) the fine-tuned PubMedBERT model from our previous study to extract miRNA-mRNA-disease associations from the entire PubMed and (ii) KinderMiner, a literature mining web-based tool, for extracting miRNA-drug interactions from the entire PubMed. We built the network to visualize novel associations among miRNA, mRNA, disease, and drug. Our study focuses on six chronic diseases: COPD, AD, stroke, type 2 diabetes mellitus (T2DM), chronic liver disease and cancer. Integrating mRNA-miRNA-drug-disease relationships is essential for understanding complex regulatory mechanisms and developing personalized medicine. Our approach advances the current state-of-the-art for drug repurposing through a novel four entity network that connects the biological and therapeutic components to identify promising agents.

## Methods

### Overview

Our proposed pipeline *(i)* extracts the miRNA-mRNA relations for chronic obstructive pulmonary disease (COPD), Alzheimer's disease (AD), stroke, type 2 diabetes mellitus (T2DM), and chronic liver disease, from PubMed using a fine-tuned PubMedBERT model from our recent study [13], *(ii)* summarizes the potential drugs for each extracted miRNA using KinderMinder, a simple literature mining tool for relation extraction [14], (iii) validates the extracted drugs using MIMIC IV database [15], (iv) visualizes the relations among miRNA, mRNA, disease, and drug by constructing three networks. Figure 1 shows the workflow.

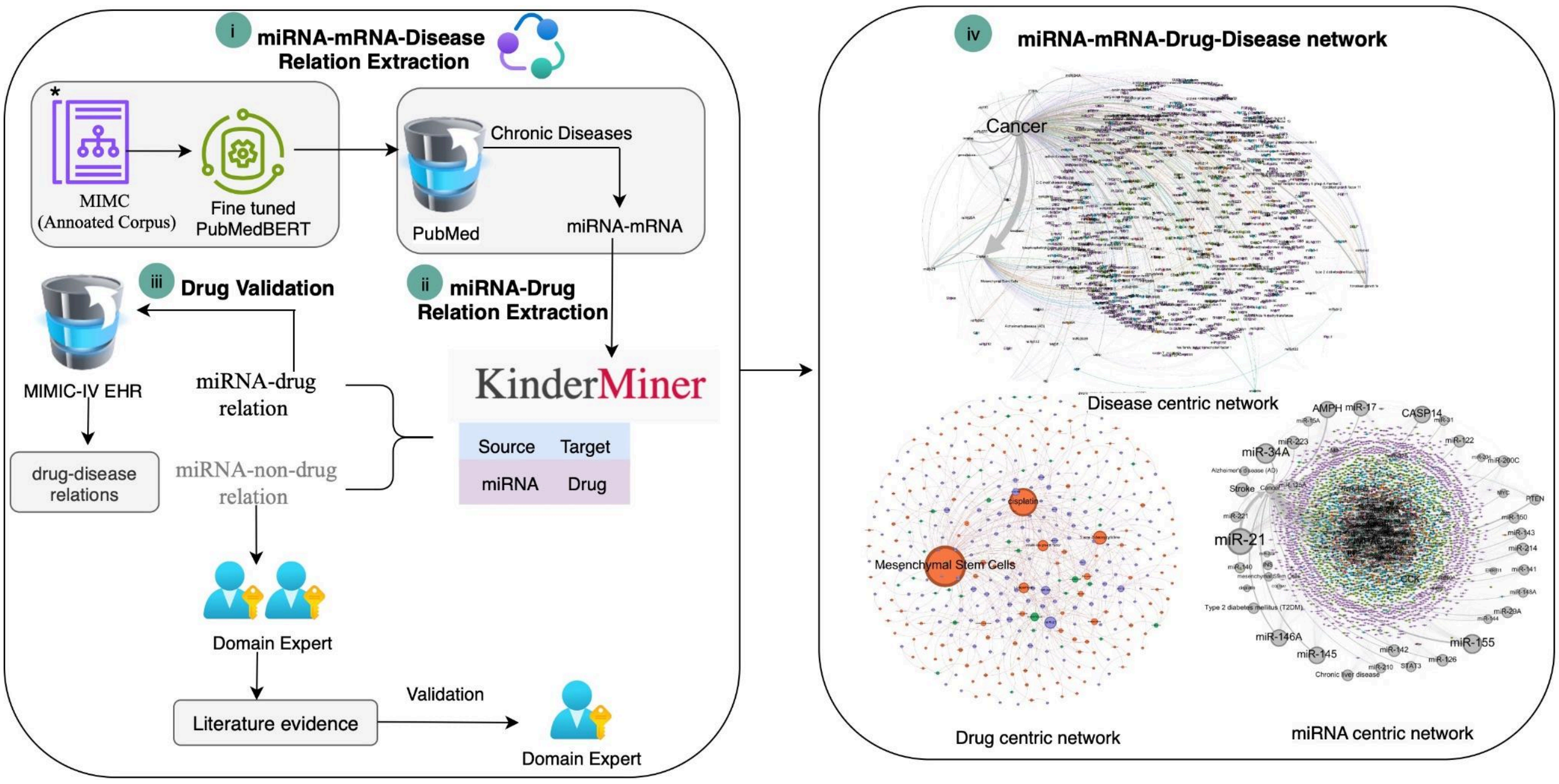

**Figure 1: Integrated framework for extracting and validating miRNA–mRNA–disease and miRNA-drug relations.**

### LLM pipeline for miRNA-mRNA-disease relation extraction

In a recent study, we evaluated the performance of machine learning (ML), deep learning-based transformer (DLT) models, and large language models (LLMs) on extracting the relations between miRNA-mRNA from PubMed abstracts [13]. We included seven ML algorithms, support vector machine (SVM), logical regression, k-nearest neighbor (KNN), decision tree, random forest, XGBoost, and LightGBM. Likewise, we included three DLT models, ClinicalBERT, BioBERT, and PubMedBERT, and four LLMs, GPT-3.5, GPT-4, Claude 2, and Llama 2. We performed zero-shot and three-shot fine-tuning for all four LLMs. We developed an annotated corpora called "miRNA–mRNA Interaction Corpus" (MMIC) for evaluating the performance. Our study revealed the best performance by PubMedBERT with an F1-score of 0.783 on MMIC.

In the current study, we applied the best performing fine-tuned PubMedBERT model on the preprocessed AI-ready input sentences from PubMed to retrieve miRNA-mRNA relations for six chronic diseases namely COPD, AD, stroke, T2DM, chronic liver disease, and cancer. Our preprocessing pipeline includes *(1)* sentence segmentation and *(2)* miRNA and mRNA recognition using scispaCy, (https://spacy.io/universe/project/scispacy, accessed on February 21 2025). The approach retrieved 577,064 sentences from 51,295 PubMed abstracts for all six diseases, COPD, AD, stroke, T2DM, chronic liver disease, and cancer.

**Literature Mining for miRNA-drug relation extraction**

miRNA-drug interactions are of significant interest in biomedical research due to their potential to regulate gene expression, influence therapeutic outcomes, and enhance personalized medicine. For each miRNA from the mRNA-miRNA-disease relationships, we retrieved a list of potential drugs from PubMed abstracts using KinderMiner, a web-based literature mining system for extracting the relations between two terms (i.e. source and target) mentioned in the PubMed abstracts [14]. We kept the default settings and used KinderMiner's drugs list dictionary as the target term. Each miRNA is considered as the source term. KinderMiner ranked the potential drugs for each miRNA determined using the Fisher Exact Test. Our attempt to retrieve miRNA-drug relations using KinderMiner identified a fewer number of miRNA - non-drug relations.

**Validating miRNA-non-drug relations**

miRNA-non-drug relations retrieved by KinderMiner were analyzed by two experts with background in pharmacology. A detailed guideline and a file with the list of miRNA-non-drug pairs were given to the experts. The file included miRNA in the first column, non-drug entity in the second column, and non-drug entity type in the third column (e.g. amino acid for alanine, a non-drug entity). The experts were asked to find at least one literature evidence from PubMed that validates the miRNA-non-drug interactions. The literature evidence from two experts (expert-1 and expert-2) for a specific miRNA-non-drug may not be the same due to the large volume of published literature and possible availability of multiple evidence. In order to validate the evidence cited by both experts, we asked them to present the related sentence(s) and section (title and/or abstract). A new expert (expert-3) with a background in pharmacology validated the literature evidence provided by expert-1 and expert-2. In addition, expert-3 obtained the agreement and disagreement statistics between expert-1 and expert-2. When expert-1 and expert-2 provided literature evidence for a miRNA-non-drug interaction, it was considered as an agreement between the experts. When one expert provides literature evidence and another does not provide the literature evidence, it was considered as a disagreement between expert-1 and expert-2.

**MIMIC-IV for validating drugs**

miRNAs play an important role in several biological processes including mRNA expression, cell differentiation, and pathogenesis of diseases [13]. Thus, the drugs identified by KinderMiner may be mapped to the diseases under study through miRNA (e.g. drug - miRNA - disease). Using the Medical Information Mart for Intensive Care (MIMIC)-IV [15,16], a large collection of de-identified patients' data, we evaluated the usefulness of the identified drugs in patients. We performed an automated search on the prescription table of MIMIC-IV to retrieve the prescription frequency and patient count prescribed with a specific drug. Here, prescription frequency refers to the number of times a drug is mentioned in the entire table and patients count is the unique number of patients prescribed with the drug under consideration. Both prescription frequency and patient counts are restricted to six diseases under study, COPD, AD, stroke, T2DM, chronic liver disease, and cancer.

We utilized the MIMIC-IV dataset to validate the drugs retrieved through miRNA-drug relation. We extracted drug-related information from the medication table, which records prescriptions. This table contains a vast collection of 15,416,708 records, corresponding to 158,421 patients and covering 9,613 different prescribed drugs. Our analysis specifically focused on identifying the drugs commonly prescribed to patients within this dataset. Through our study, we identified 27 drugs that were consistently prescribed across different patients. These drugs were then analyzed further to understand their prescription patterns, including their frequency of use and the patient count for each.

**miRNA, mRNA, disease, drug relation visualization**

We generated three networks, disease-centric network, drug-centric network, and miRNA-centric networks using Gephi, a network visualization and analysis tool (https://gephi.org/). The inputs were the extracted miRNA-mRNA-disease triplets using PubMedBERT and miRNA-drug pairs using KinderMiner.  was input into the Gephi network visualization and analysis tool. The networks uncovered many novel associations and provided clues about shared molecular mechanisms, therapeutic potential, and possible crossovers in drug repurposing. The overall findings have significant implications for understanding disease mechanisms at the molecular level, identifying potential therapeutic targets, and exploring new drug-disease connections that could inform future research and treatment strategies.

For creating and analyzing the three networks, we first loaded the main graph into Gephi for initial visualization and analysis. The Filter panel in Gephi was then used to define the subset of nodes and edges of interest. We filtered nodes based on attributes such as degree (the number of connections a node has) and modularity class (community detection). Specifically, we navigated to Filters -> Attributes and applied appropriate filters, such as Range or Topology, to isolate nodes and edges based on properties like degree or community membership. Once the desired filter was selected, it was dragged to the Queries area, and filter values were adjusted as needed. By clicking Filter, we applied the filter to visually isolate the nodes and edges that belonged to the specific subnetwork. The specific sub-networks were generated by selecting mRNAs with large connections with at least two diseases or drugs.

For a deeper analysis using networks, we ranked the nodes by their betweenness centrality, which measures how often a node appears on the shortest paths between other nodes. Betweenness centrality is a key indicator of a node's influence within the network. Nodes with higher betweenness centrality are considered more critical for maintaining connections between different parts of the network. In Gephi, nodes were ranked by betweenness centrality, and their size was adjusted accordingly, where larger nodes indicate higher betweenness centrality and greater connectivity. In general, nodes with a larger size represent those that have a larger number of connections or play a more influential role in bridging different parts of the network.

## Results

### Extracted miRNA-mRNA-disease relations

The dataset included 1,169 PubMed abstracts with at least one miRNA for COPD, 4,257 for AD, 5,027 for stroke, 3,117 for T2DM, 2,830 for chronic liver disease, and 34,895 for cancer. Our approach filtered a huge volume of PubMed abstracts without miRNA mentions that are not required for the current study. A total of 311 sentences with miRNA and mRNA mentions for COPD, 452 for AD, 541 for stroke, 1,112 for T2DM, 1,048 for chronic liver disease, and 42,993 for cancer were used as input for the fine-tuned PubMedBERT model (Table 1). The goal was to classify these sentences with / without miRNA-mRNA association. Our approach identified 130 sentences miRNA-mRNA relations for COPD, 296 relations for AD, 446 relations for stroke, 356 relations for T2DM, 326 relations for chronic liver disease, and 28,068 relations for cancer (Table 1, Supplementary Table 1). We normalized miRNAs to a uniform pattern, i.e., miR-NUMBER-NUMBER or miR-NUMBERalphabet and mapped mRNA to the official gene symbol using STRING database, a popular online resource for protein-protein interactions information [17]. For sentences with more than one miRNA or mRNA, we generated templates to represent only one miRNA, mRNA or miRNA-mRNA relations per template. The sentences with miRNA-mRNA associations for all six diseases under study included 807 miRNAs (normalized to a uniform pattern, i.e., miR-NUMBER-NUMBER or miR-NUMBERalphabet) and 2,684 mRNAs (normalized to official gene symbol) (Supplementary Table 2).

Table 1. miRNA-mRNA-disease relations

| Disease | PubMed abstracts with miRNA(s) | Sentences | Sentences with miRNA and mRNA mentions (%) | Sentences with miRNA-mRNA relations (%) |
|---|---|---|---|---|
| COPD | 1,169 | 8,183 | 311 (3.8%) | 130 (1.6%) |
| AD | 4,257 | 46,827 | 452 (1%) | 296 (0.6%) |
| Stroke | 5,027 | 60,324 | 541 (0.9%) | 446 (0.7%) |
| T2DM | 3,117 | 46,755 | 1,112 (2.4%) | 356 (0.8%) |
| Chronic liver disease | 2,830 | 31,130 | 1,048 (3.4%) | 326 (1.1%) |
| Cancer | 34,895 | 383,845 | 42,993 (11.2%) | 28,068 (7.3%) |

**Extracted miRNA-drug relations**

Table 2 shows miRNA-drug relations retrieved by KinderMiner. Among 807 miRNAs associated with six diseases under study, 181 miRNAs (22.4%) show significant relation with 153 drugs (Supplementary Table 3). The total number of relations predicted by KinderMinder is 637. Manual review of retrieved relations by two domain experts with background in pharmacology revealed 595 relations between miRNA-drug (Supplementary Table 3) and 42 relations between miRNA-non-drug (Supplementary Table 4). Manual analysis by the experts also revealed that 73 miRNAs are connected with cisplatin, a powerful platinum-based chemotherapy drug used to treat various cancers, including bladder, ovarian, and testicular cancers. Interestingly, 14 (out of 73) miRNAs are linked only to cisplatin. miR-21 is significantly linked with a maximum of 30 different drugs: (-)-rapamycin, alkaline Phosphatase, angiotensin ii, apocynin, arsenite, berberine, bleomycin, celastrol, cisplatin, curcumin, diindolylmethane, dovitinib, doxorubicin, fluorouracil, gemcitabine, hemin, ly294002, mesenchymal stem cells, metformin, molybdenum, paclitaxel, propidium, resveratrol, sb431542, silibinin, sorafenib, streptozotocin, temozolomide, u0126, and wp1066. Interestingly, certain miRNAs are significantly linked with only one drug (e.g. miR-1229 and actinomycin, miR-125B2 and aminoflavone). Our analysis also revealed that 626 miRNAs have no relations with the drugs. miRNA-drug relation is comparatively a new approach when compared to mRNA-drug relations. In specific, miRNA mention (a total of 807) is much lower than mRNA mentions (a total of 2,684) in PubMed abstracts (with at least one miRNA mention) related to six diseases under study.

Table 2. Significant miRNA-drug relations predicted by KinderMiner

| **Disease** | **miRNA (Source)** | **miRNA-drug relations (%)** |
|---|---|---|
| COPD | 23 | 170 (28.6%) |
| AD | 45 | 202 (33.9%) |
| Stroke | 62 | 259 (43.5%) |
| T2DM | 51 | 281 (47.2%) |
| Chronic liver disease | 44 | 268 (45.0%) |
| Cancer | 802 | 594 (99.8%) |

**Literature evidence for miRNA-non-drug relations**

Two domain experts with pharmacology background defined the category for each non-drug (e.g. amino acid as the category for alanine, a non-drug retrieved by KinderMiner) and validated 42 miRNA-non-drug relations by identifying at least one literature evidence that confirms the relation. The experts worked independently. A third domain expert with a background in pharmacology verified the validations and summarized the experts' opinion (Supplementary Table 4). Most of the non-drug entities retrieved by KinderMiner are amino acids (e.g. alanine, phosphocreatine), proteins (e.g. fibroblast growth factor, tumor necrosis factor receptor), and enzymes (e.g. lipase, lactase). Domain expert 1 submitted literature evidence for all 42 miRNA-non-drug relations. The literature evidence for two miRNA-non-drug relations, miR-216A-lipase and miR-217-lipase, show indirect relation via pancreatic injury. Domain expert 2 submitted literature evidence for all 42 miRNA-non-drug relations. There was no evidence showing indirect relations.

**Drug mentions in MIMIC-IV**

Among 75 drugs from the miRNA-drug relations, 27 drugs are in the MIMIC-IV database (Table 3). The drug prescription frequency ranges from one to 176. The number of patients who received the drugs ranges from one to 85. For diagnoses (i.e. disease), we report only for the diseases under study. The diagnoses data shows zero to five diseases under study for each drug. While propranolol, pilocarpine, and inositol are not prescribed to patients diagnosed with the diseases under study, lenalidomide is prescribed to patients diagnosed with COPD, AD, T2DM, chronic liver disease, or cancer. We also observed that the prescription frequency and patient count are very low for propranolol, pilocarpine, and inositol when compared to lenalidomide. Among 27 drugs in MIMIC-IV, 20 drugs (74.1%) are prescribed to patients diagnosed with T2DM or cancer. Likewise, 17 drugs (63%) are prescribed to patients diagnosed with chronic liver disease, 9 drugs (33.3%) are prescribed to patients diagnosed with COPD, and 2 drugs (7.4%) are prescribed to patients diagnosed with AD. Interestingly, stroke is not treated with any of the drugs predicted through miRNA-drug relation. For chronic liver disease and cancer, we consider any subtypes (e.g.

chronic liver cirrhosis, neoplasm of liver, neoplasm of breast, neoplasm of bladder). Our approach to validate the repurposed drugs via miRNA interaction using MIMIC-IV shows promising results. Thus, repurposing drugs for a disease via miRNA interaction is a novel approach.

Table 3. Drug prescription frequency, patients and diagnoses from MIMIC-IV

| Drug | Prescription Frequency | Patients | Diagnoses* | | | | | |
|---|---|---|---|---|---|---|---|---|
| | | | COPD | AD | Stroke | T2DM | Chronic liver disease | Cancer |
| venetoclax | 176 | 43 | ✓ | ✗ | ✗ | ✓ | ✓ | ✓ |
| melatonin | 131 | 85 | ✓ | ✓ | ✗ | ✓ | ✗ | ✓ |
| temozolomide | 71 | 26 | ✗ | ✗ | ✗ | ✓ | ✓ | ✓ |
| enzalutamide | 63 | 34 | ✓ | ✗ | ✗ | ✓ | ✓ | ✓ |
| lenalidomide | 50 | 32 | ✓ | ✓ | ✗ | ✓ | ✓ | ✓ |
| abiraterone | 40 | 24 | ✓ | ✗ | ✗ | ✓ | ✓ | ✓ |
| ticagrelor | 30 | 13 | ✗ | ✗ | ✗ | ✓ | ✓ | ✓ |
| sorafenib | 24 | 15 | ✗ | ✗ | ✗ | ✓ | ✓ | ✓ |
| simvastatin | 23 | 17 | ✗ | ✗ | ✗ | ✓ | ✗ | ✓ |
| lidocaine | 22 | 22 | ✓ | ✗ | ✗ | ✓ | ✓ | ✓ |
| vemurafenib | 19 | 9 | ✗ | ✗ | ✗ | ✓ | ✓ | ✓ |
| sofosbuvir | 17 | 10 | ✗ | ✗ | ✗ | ✓ | ✓ | ✓ |
| erlotinib | 16 | 13 | ✗ | ✗ | ✗ | ✓ | ✓ | ✓ |
| estradiol | 16 | 14 | ✓ | ✗ | ✗ | ✗ | ✓ | ✓ |
| sunitinib | 15 | 9 | ✗ | ✗ | ✗ | ✗ | ✓ | ✓ |
| ribavirin | 15 | 8 | ✗ | ✗ | ✗ | ✓ | ✓ | ✓ |
| olaparib | 10 | 5 | ✗ | ✗ | ✗ | ✓ | ✗ | ✓ |
| fluorouracil | 8 | 6 | ✗ | ✗ | ✗ | ✓ | ✗ | ✓ |
| telaprevir | 8 | 4 | ✗ | ✗ | ✗ | ✗ | ✓ | ✗ |
| daclatasvir | 6 | 4 | ✓ | ✗ | ✗ | ✓ | ✓ | ✗ |
| metformin | 6 | 3 | ✗ | ✗ | ✗ | ✓ | ✓ | ✗ |
| medroxyproge sterone | 4 | 3 | ✗ | ✗ | ✗ | ✓ | ✗ | ✓ |
| propranolol | 3 | 2 | ✗ | ✗ | ✗ | ✗ | ✗ | ✗ |
| gefitinib | 3 | 3 | ✓ | ✗ | ✗ | ✗ | ✗ | ✓ |
| pilocarpine | 2 | 1 | ✗ | ✗ | ✗ | ✗ | ✗ | ✗ |
| clopidogrel | 2 | 2 | ✗ | ✗ | ✗ | ✓ | ✗ | ✗ |
| inositol | 1 | 1 | ✗ | ✗ | ✗ | ✗ | ✗ | ✗ |
| **Total drugs** | **-** | **-** | **9** | **2** | **0** | **20** | **17** | **20** |

*Diseases under study ONLY

The analysis based on the prescription frequency provides key insights into drug usage patterns and their associated conditions (Table 3). Venetoclax, prescribed 176 times to 43 patients, is primarily used for treating chronic lymphocytic leukemia, small lymphocytic lymphoma, and acute myeloid leukemia, indicating its role in managing various blood cancers. Melatonin, with 131 prescriptions for 85 patients, is mainly prescribed for circadian rhythm sleep disorders, reflecting its wide use in sleep management. Temozolomide, prescribed 71 times for 26 patients, targets brain cancers such as glioblastoma and refractory anaplastic astrocytoma. Enzalutamide, used 63 times in 34 patients, is crucial for treating different stages of prostate cancer, including castration-resistant and HRR gene-mutated cases. Lenalidomide, which appears 50 times for 32 patients, addresses a range of hematologic conditions like multiple myeloma, transfusion-dependent anemia, and several types of lymphoma. Table 3 further highlights the usage of other drugs such as abiraterone, frequently used for advanced prostate cancer, and ticagrelor, lidocaine, and sorafenib, among others, showing their importance in treating diverse conditions such as cardiovascular issues, pain management, and cancer. The prescription frequency highlights the usage trends of these drugs in the real world, reflecting both common and niche applications. The patient counts also indicate how widely these drugs are used across the patient population. Drugs like venetoclax and melatonin are more frequently prescribed and thus cater to a larger patient base, whereas specialized drugs like vemurafenib and sunitinib are used less frequently, typically in the treatment of rare or advanced cancers.

In addition to MIMIC-IV, we validated 27 drugs using DrugBank (https://go.drugbank.com/) and Comparative Toxicogenomics Database (CTD) (https://ctdbase.org/). For every drug, we manually retrieved the indications (i.e., diseases) from DrugBank. For CTD, we considered the diseases marked as “Direct Evidence” and “Therapeutic”, and removed the duplicates. Our approach identified one to seven diseases (all) for 27 drugs from DrugBank and one to 131 diseases (all) for 27 drugs from CTD (Supplementary Table 5). Most drugs are meant for treating one or more diseases under study. Ticagrelor, simvastatin, and clopidogrel are used for treating stroke. Simvastatin is used for treating COPD. Melatonin is used for treating Alzheimer’s disease. Metformin is used for treating T2DM. For chronic liver diseases and cancer, we see multiple specific liver diseases (e.g. liver cirrhosis, fatty liver disease) or cancer subtypes (e.g. neoplasm of breast, leukemia, melanoma) associated with 27 drugs in both DrugBank and CTD. The finding reveals that drug repurposing through miRNA-drug relation is a promising approach.

**miRNA-centric, drug-centric, and disease-centric network analysis**

**miRNA-centric network:** Figure 2 shows a miRNA-centric network connecting miRNA, mRNA, disease, and drug related to six diseases under study. A total of 3,497 nodes and 16,417 edges were organized as a directed graph. Despite its size, the graph exhibits a low density (0.001) with an average path length of 1.386, reflecting a sparse yet highly connected small-world structure. Importantly, the presence of a single connected component highlights that all entities are interlinked within one unified system. The network shows a heterogeneous topology with an average degree of 4.7 and a substantially higher average weighted degree of 28.8, pointing to a subset of nodes acting as major hubs with broad connectivity across biological domains.

Several hubs stand out as biologically and clinically significant, including miR-21, miR-155, miR-34A, miR-146A, and miR-145, which are connected to multiple diseases such as AD, stroke, T2DM, chronic liver disease, and cancer. mRNAs such as CASP14, STAT3, PTEN, and MYC, along with therapeutic agents such as cisplatin and mesenchymal stem cells, also emerge as key connectors that bridge different clusters of the network. The average clustering coefficient of 0.307 reflects strong local interconnectivity, yet the low modularity value (0.008) indicates that the network is not cleanly partitioned, with overlapping modules centered around major miRNA hubs. These findings underscore the role of specific miRNAs as master regulators that integrate multiple pathways, making them particularly promising for mechanistic insights and translational applications.

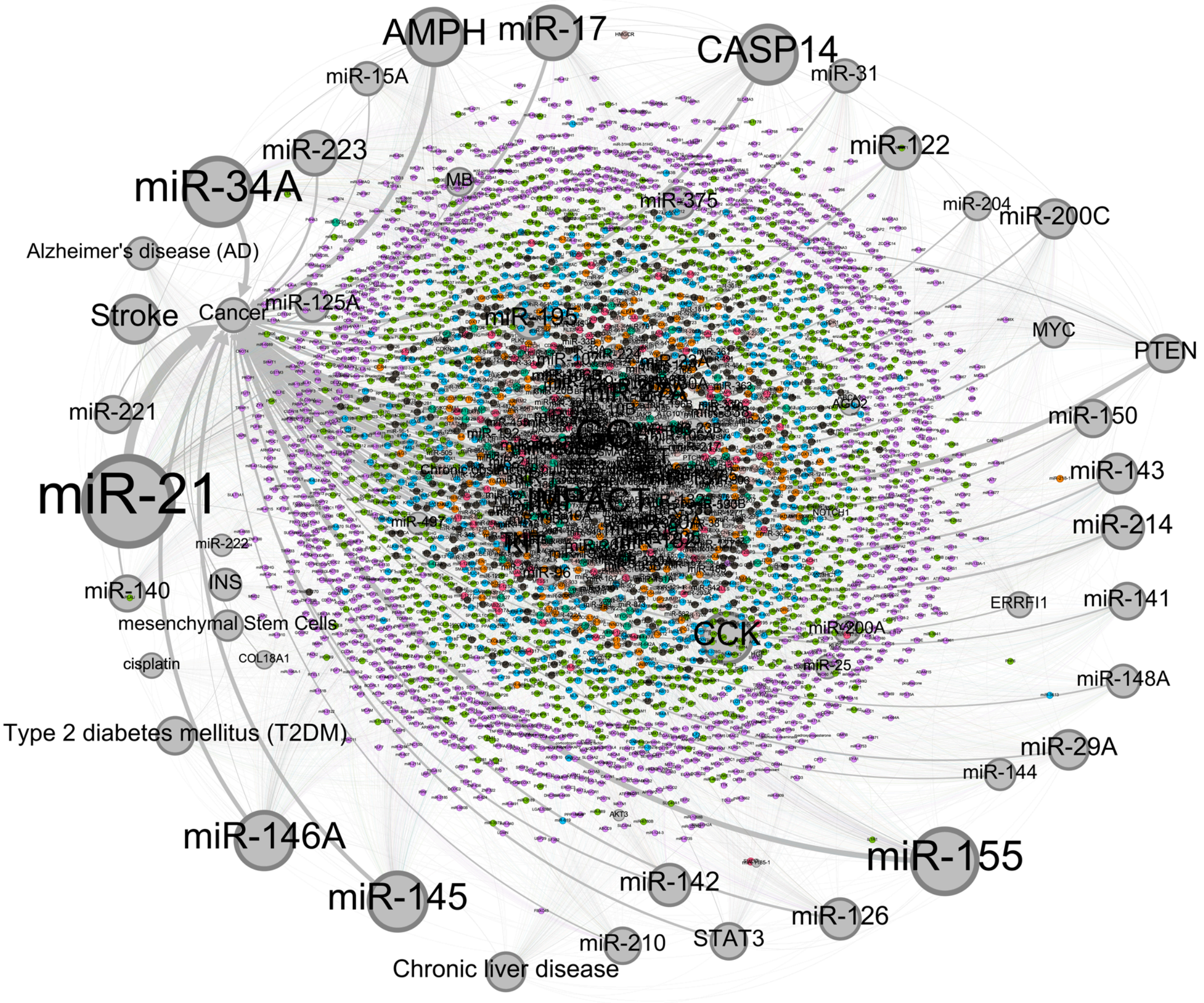

Figure 2. miRNA-mRNA-disease-drug network highlighting significant miRNAs

**Drug-centric network:** Figure 3 shows a drug-centric network, where drugs are placed at the center of their interaction profiles with miRNAs, mRNAs, and diseases. The node size is proportional to connectivity, highlighting key drugs that act as major hubs within the network. Notably, mesenchymal stem cells and cisplatin dominate as the largest hubs, indicating their widespread interactions with multiple miRNAs and mRNAs, as well as diverse disease contexts. These nodes function as bridging entities, linking several biological modules together, which reflects their extensive therapeutic relevance. Other interesting nodes include rapamycin, insulin-like growth factor, and 5-aza-2′-deoxycytidine, which also show considerable connectivity and suggest roles in regulating multiple pathways. Surrounding clusters represent smaller drug–miRNA–mRNA interaction modules, but they are interconnected through these major hubs.

The presence of miRNAs such as miR-21 and miR-34 in proximity to key drug nodes highlights how these small RNAs mediate drug responses and therapeutic effects. Overall, this drug-centric network emphasizes the central regulatory influence of a few key drugs/biologics and their ability to rewire miRNA–mRNA-disease relations, offering insight into potential drug repurposing opportunities and combination therapies.

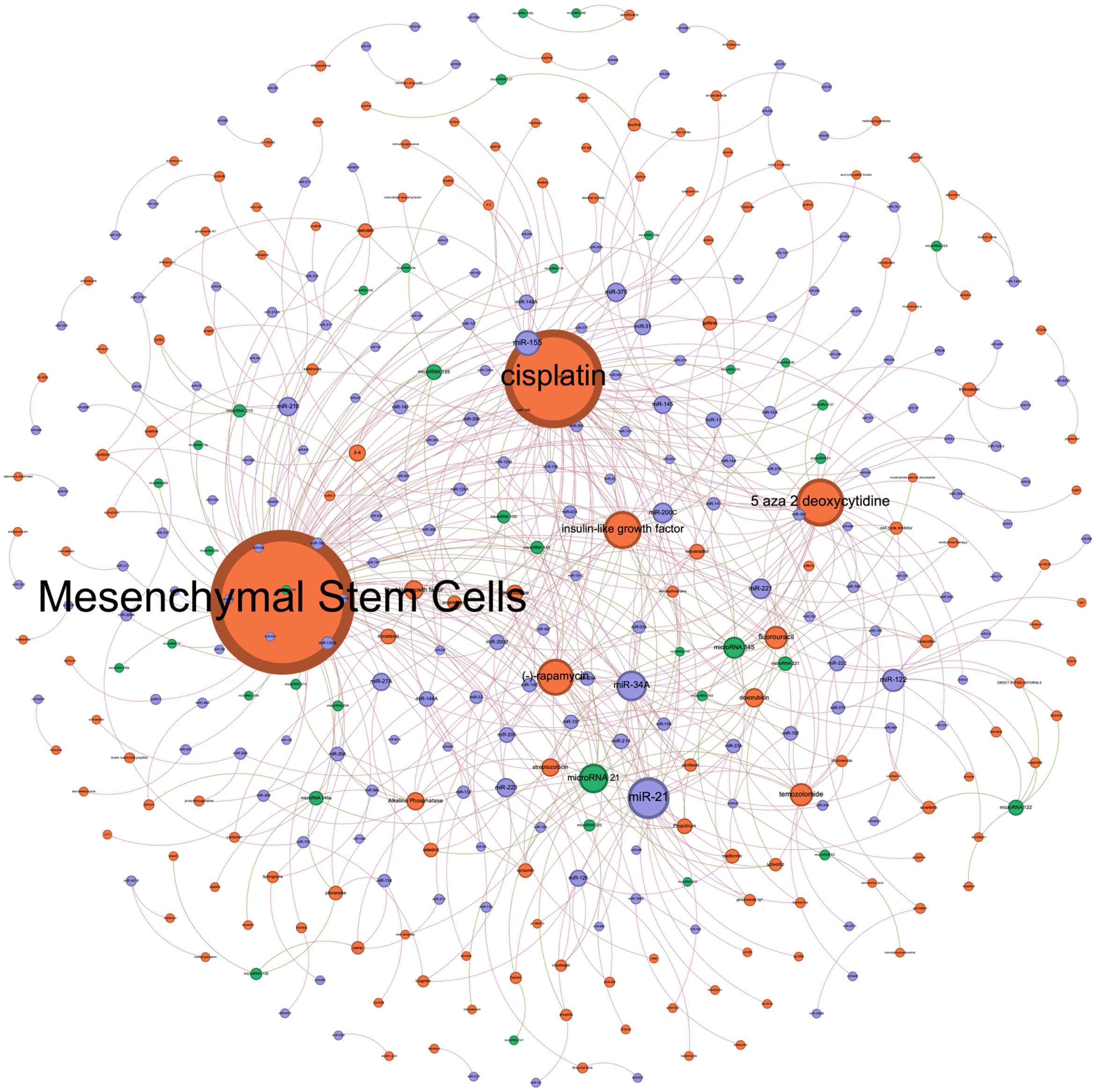

Figure 3. miRNA-mRNA-disease-drug network highlighting significant drugs

**Disease-centric network:** Figure 4 highlights the central role of cancer as a major hub connecting to a wide range of miRNAs, mRNAs, and drugs. The node size corresponds to connectivity, with cancer dominating the network structure and linking to numerous regulatory molecules. Prominent miRNAs such as miR-21, miR-34A, miR-146A, and miR-155, along with therapeutic agents like cisplatin and mesenchymal stem cells, appear closely connected to cancer, reflecting their well-established involvement in oncogenic pathways. The dense web of interactions indicates that cancer serves as an integration point for diverse molecular mechanisms, with extensive overlap across other diseases including COPD, AD, stroke, and T2DM. The network also reveals how cancer-associated miRNAs and mRNAs form highly interconnected clusters that extend into non-cancer disease modules, suggesting shared regulatory pathways and potential cross-disease therapeutic targets. The presence of strong links between cancer and nodes such as PTEN, STAT3, and MYC underscores the relevance of well-known tumor suppressors and oncogenes. Meanwhile, peripheral clusters represent disease-specific subnetworks that remain tethered to cancer through hub miRNAs and drugs, illustrating the pleiotropic role of miRNA regulation in linking oncological and non-oncological diseases.

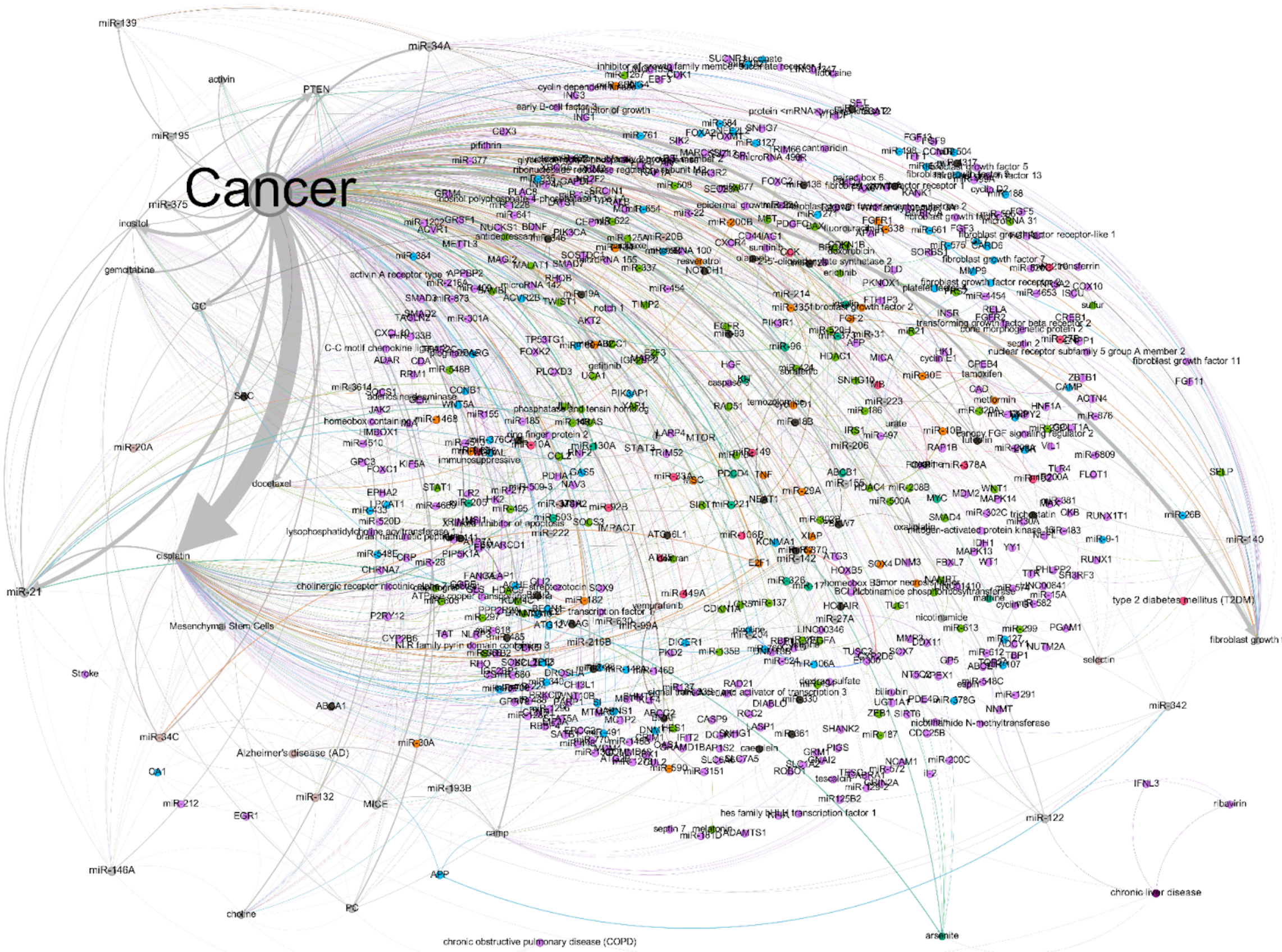

Figure 4. ImiRNA-mRNA-disease-drug network highlighting significant diseases

The construction of these networks are grounded in relations mined from PubMed sentences, focusing on six major diseases of high global burden— COPD, AD, stroke, T2DM, chronic liver disease, and cancer. These diseases are widely studied in biomedical literature, and the extracted associations naturally highlight well-characterized regulators such as miR-21, miR-155, miR-34A, miR-146A, and miR-145, which frequently appear in published contexts linking them to multiple disease pathways. The emergence of these miRNAs as major hubs reflects not only their biological importance but also the literature-driven connectivity that underscores their role as central regulators across different pathological processes.

Likewise, the prominence of mRNAs (CASP14, STAT3, PTEN, MYC) and drugs/biologics such as cisplatin and mesenchymal stem cells is consistent with their extensive representation in PubMed abstracts describing mechanistic studies and therapeutic interventions. The overlap across disease modules, as reflected by the low modularity score, can be interpreted as a consequence of shared molecular mechanisms that are repeatedly reported in diverse disease contexts. This literature-derived integration reinforces the idea that these diseases do not exist in isolation but rather share common regulatory circuits mediated by key miRNAs and mRNAs. Therefore, the network not only captures biological reality but also represents the collective knowledge landscape curated from scientific publications, positioning it as a valuable resource for identifying cross-disease therapeutic targets and opportunities for drug repurposing.

**Discussion**

MicroRNA (miRNA) interactions with drugs are of significant interest in biomedical research due to their potential to regulate gene expression, influence therapeutic outcomes, and enhance personalized medicine. The study of miRNA-drug interactions is pivotal for advancing therapeutic interventions, improving drug efficacy and safety, and fostering the development of personalized medicine approaches.

miRNA-drug relations in PubMed abstracts indicate the importance of miRNA in influencing drug interactions. miR-21 shows the highest connectivity, being linked with 30 different drugs, including chemotherapeutics such as cisplatin, gemcitabine, and doxorubicin. This finding supports the established role of miR-21 in cancer biology, where it controls the pathways related to cell survival, programmed cell death, and immune system functions [18,19,20]. On the other hand, miR-107 is linked to two drugs (cisplatin and conjugated linoleic acid), suggesting the possibility of a more specific therapeutic role. In specific, miR-107 plays a critical role in overcoming cisplatin resistance in non-small cell lung cancer (NSCLC) by targeting CDK8, which is linked to drug resistance. The significantly lower expression of CDK8 in NSCLC cells compared to the normal cells suggests the possibility of miR-107 serving as a predictive marker for chemotherapy response [21]. The study also validated these miRNA-drug associations using the MIMIC-IV database, analyzing prescription frequencies and patient counts to provide real-world evidence for the therapeutic application of identified drugs. This integration of clinical data ensures that the miRNA-drug associations identified are not only theoretically plausible but also clinically relevant. Through network visualization using Gephi, the study identified disease-centric, miRNA-centric, and drug-centric sub-networks, revealing the shared molecular pathways and potential drug repurposing opportunities for diseases such as COPD, AD, stroke, T2DM, chronic liver disease and cancer. The centrality measures and modularity detection highlighted key miRNAs, such as miR-21, miR-146A, and miR-342, that serve as central players in these networks. This approach offers a comprehensive view of the interactions and identifies potential targets for therapeutic interventions.

The study emphasizes the involvement of miRNAs across various diseases and biological processes. For example, miR-21 is highly upregulated in numerous pathological conditions including cancer, cardiovascular diseases, pulmonary fibrosis, and central nervous system (CNS) disorders [22,23]. It targets several tumor suppressor genes (e.g., PTEN, PDCD4) and influences the TGF-β and NF-κB pathways, playing a critical role in inflammation, apoptosis, and immune regulation [24,25,26,27]. In cardiovascular diseases, miR-21 plays a significant role in various conditions, including heart failure, myocardial infarction, and the injuries that occur when blood flow is restored after a heart attack [28,29,30,31]. This highlights the influence of miR-21 in heart health and disease. In cancer, miR-21 promotes tumor growth, invasion, and metastasis by downregulating tumor suppressor genes such as PTEN and PDCD4. Its role in chemotherapy resistance, particularly with cisplatin, is evident in several studies where miR-21 is shown to inhibit apoptosis pathways, thus contributing to drug resistance. This effect is notably observed in gastric, ovarian, and non-small cell lung cancers [32,33,34]. Additionally, the role of miR-21 extends to non-cancer conditions such as asthma, fibrosis, and CNS disorders like Alzheimer's disease and ischemic stroke [35,36,37,38,39]. It regulates inflammation through pathways such as TLR4/NF-κB and MAPK and influences neurological outcomes by modulating apoptosis and protecting the blood-brain barrier [40,41]. The diverse functionality of miR-21 demonstrates its importance as a therapeutic target, although its broad involvement across multiple diseases limits its specificity as a biomarker.

Similar to miR-21, miR-146A is implicated in both cancer and chronic diseases such as COPD (Chen et al. 2018). It regulates inflammation and immune responses by impacting pathways like NF-κB, MEK-1/2, and JNK-½ [42]. The study points out miR-146A's dual role, highlighting its influence in cancer progression and inflammatory responses in COPD [43]. This indicates miR-146A as a significant target for therapeutic strategies to modulate immune and inflammatory pathways. miR-342, another miRNA highlighted in the study, is associated with cancer, Alzheimer's disease, and T2DM [44,45,46,47]. It plays a role in regulating neuroinflammation, insulin resistance, and amyloid processing, linking metabolic and neurodegenerative conditions with cancer pathways [48,49]. The shared miRNA networks, particularly those involving miR-21, miR-146A, and miR-342, present opportunities for drug repurposing (Supplementary Figures 1, 2, and 3). For instance, miRNAs involved in cancer therapy could be targeted to treat other chronic conditions such as COPD or liver disease, where similar pathways are affected. This approach could enhance the efficiency of drug development and application by leveraging existing cancer drugs for other conditions.

Limitations: The miRNA mentions in PubMed is very low when compared to mRNA mentions in PubMed. This resulted in small data for COPD, AD, stroke, T2DM, and chronic liver disease. Cancer data is the largest counting for 28,068 sentences with miRNA-mRNA relations from 34,895 PubMed abstracts. If we split this data per cancer subtypes, it is possible that many subtypes may not have the data in PubMed. Our results are more leaning towards cancer because of the uneven distribution of data among the diseases under study. In the future, we will study the role of miRNA in specific cancer subtypes for more insights.

## Conclusion

The study presents a novel approach to drug repurposing for chronic diseases by constructing miRNA-mRNA-drug-disease networks. This comprehensive method highlights the pivotal role of miRNAs in regulating gene expression and disease progression across various conditions. By integrating LLMs, text-mining techniques, network analysis, and clinical data, the research reveals significant miRNA-drug interactions and potential therapeutic targets. The findings demonstrate the value of such integrated networks for discovering drug repurposing opportunities and advancing personalized medicine, providing insights into complex disease mechanisms and therapeutic interventions. The study also shows that the analysis based on miRNAs involved in diseases such as COPD, AD, stroke, T2DM, and chronic liver disease can reveal the comorbidity connectivity with cancer through the network analysis.

**References:**

1. Friedman, R.C.; Kyle K.F.; Christopher B.B.; David P.B. Most Mammalian mRNAs Are Conserved Targets of microRNAs. *Genome Research* 2009, 19 (1), 92–105.

2. Rooij, L.A.; Dirk J.M.; Nicky T.V.; Elsken W.; Paul J.D.; Cathy B.M. The microRNA Lifecycle in Health and Cancer." *Cancers* 2022, 14 (23). https://doi.org/10.3390/cancers14235748.

3. Fischer, S.; René, H.; Armaz, A.; Kerstin, O. Unveiling the Principle of microRNA-Mediated Redundancy in Cellular Pathway Regulation." *RNA Biology* 2015, 12 (3), 238–47.

4. Jang, H.J.; Sang I.Lee. MicroRNA Expression Profiling during the Suckling-to-Weaning Transition in Pigs. *J. Animal Sci. Tech.* 2021, 63 (4), 854–63.

5. Wang, B.; Shuqiang L.; Hank H.Q.; Dipanjan C.; Yang S.; Carl D.N. Distinct Passenger Strand and mRNA Cleavage Activities of Human Argonaute Proteins." *Nat. Struc. Mole. Bio.* 2009, 16 (12), 1259–66.

6. Ritchie, W.; Megha, R.; Stephane, F.; John E.J.R. Conserved Expression Patterns Predict microRNA Targets. *PLOS Comput. Biol.* 2009, 5 (9): e1000513.

7. Otmani, K.; Philippe L. Tumor Suppressor miRNA in Cancer Cells and the Tumor Microenvironment: Mechanism of Deregulation and Clinical Implications. *Frontiers in Oncology* 2021, 11 (October):708765.

8. Li, S.; Zhixin, L.; Taolei S. The Role of microRNAs in Neurodegenerative Diseases: A Review. *Cell Biology and Toxicology* 2023, 39 (1), 53–83.

9. Winkle, M.; Sherien, M.E.; Muller, F.; George A.C. Noncoding RNA Therapeutics - Challenges and Potential Solutions. *Nature Reviews. Drug Discovery* 2021, 20 (8), 629–51.

10. Seyhan, A.A. Trials and Tribulations of MicroRNA Therapeutics." *Int. J. Mol. Sci.* 2024, 25 (3). https://doi.org/10.3390/ijms25031469.

11. Rhodes, L.V., Ashley, M.N.H.; Chris, S.; Elizabeth, C.M.; Jennifer, L.D.; Steven, E.; Seung, Y.N. The Histone Deacetylase Inhibitor Trichostatin A Alters microRNA Expression Profiles in Apoptosis-Resistant Breast Cancer Cells. *Oncology Reports* 2012, 27 (1), 10–16.

12. Beg, M.S.; Andrew, J.B.; Jasgit, S.; Mitesh, B.; Yoon-Koo Kang, J.St.; Susan, S,; Andreas, G.B.; Sinil, K.; David S.H. Phase I Study of MRX34, a Liposomal miR-34a Mimic, Administered Twice Weekly in Patients with Advanced Solid Tumors. *Investigational New Drugs* 2017, 35 (2), 180–88.

13. Bhasuran, B.; Sharanya, M.; Oviya, R.Iyyappan, Gurusamy, M.; Archana, P.; Kalpana, R. Large Language Models and Genomics for Summarizing the Role of microRNA in Regulating mRNA Expression. *Biomedicines* 2024, 12 (7). https://doi.org/10.3390/biomedicines12071535.

14. Kuusisto, F.; Daniel, N.;, John, S.; Ian, R.; Miron, L.; James, T.; David, P.; Ron, S. KinderMiner Web: A Simple Web Tool for Ranking Pairwise Associations in Biomedical Applications." *F1000Research* 2020, 9 (July),832.

15. Johnson, A.E.W.; Lucas, B.; Lu, S.; Alvin, G.; Ayad, S.; Steven, H.; Tom, J.P. MIMIC-IV, a Freely Accessible Electronic Health Record Dataset. *Scientific Data* 2023, 10 (1), 1–9.

16. Johnson, A.; Lucas, B.; Tom, P.; Brian, G.; Benjamin, M.; Steven, H.; Leo, A.C.; Roger, M. MIMIC-IV. 2024, https://doi.org/10.13026/kpb9-mt58.

17. Szklarczyk, D.; Rebecca, K.; Mikaela, K.; Katerina, N.; Farrokh, M.; Radja, H.; Annika, L.G. The STRING Database in 2023: Protein-Protein Association Networks and Functional Enrichment Analyses for Any Sequenced Genome of Interest. *Nucleic Acids Research* 2023, 51 (D1), D638–46.

18. Geretto, M.; Alessandra, P.; Camillo, R.; Dinara, Z.; Rakhmet, B.; Alberto, I. Resistance to Cancer Chemotherapeutic Drugs Is Determined by Pivotal microRNA Regulators. *American J. Cancer Res.* 2017, 7 (6), 1350–71.

19. Bautista-Sánchez, D.; Cristian, A.C.; Abraham, P.T.; Inti Alberto De La Rosa-Velázquez, Rodrigo González-Barrios, Laura Contreras-Espinosa, Rogelio Montiel-Manríquez, et al. 2020. "The Promising Role of miR-21 as a Cancer Biomarker and Its Importance in RNA-Based Therapeutics." *Molecular Therapy. Nucleic Acids* 20 (June):409–20.

20. Mizielska, A.; Iga, D.; Radosław, S.; Małgorzata, C.; Małgorzata, D.; Jan, Ś.; Izabela, K. Doxorubicin and Cisplatin Modulate miR-21, miR-106, miR-126, miR-155 and miR-199 Levels in MCF7, MDA-MB-231 and SK-BR-3 Cells That Makes Them Potential Elements of the DNA-Damaging Drug Treatment Response Monitoring in Breast Cancer Cells—A Preliminary Study. *Genes* 2023, 14 (3), 702.

21. Zhang, L.; Wei, O.; Qianwen, L.; Ning, L.; Li, L.; Siyu, W. Pemetrexed plus Carboplatin as Adjuvant Chemotherapy in Patients with Curative Resected Non-Squamous Non-Small Cell Lung Cancer. *Thoracic Cancer* 2014, 5 (1), 50–56.

22. Kumarswamy, R.; Ingo, V.; Thomas, T. Regulation and Function of miRNA-21 in Health and Disease. *RNA Biology* 2011, 8 (5), 706–13.

23. Bai, X.; Zhigang, B. MicroRNA-21 Is a Versatile Regulator and Potential Treatment Target in Central Nervous System Disorders. *Front. Mol. Neurosci.* 2022, 15 (January):842288.

24. Buscaglia, L.E.B.; Yong, Li. Apoptosis and the Target Genes of microRNA-21. *Chinese J. Cancer* 2011, 30 (6), 371–80.

25. Ben, H.S.; Khadija, E.B. Interplay between Signaling Pathways and Tumor Microenvironment Components: A Paradoxical Role in Colorectal Cancer" *Int. J. Mol. Sci.* 2023, 24(6). https://doi.org/10.3390/ijms24065600.

26. Prasad, M.D.; Hamsa, M.F.; Mohmed, I.K. An Update on the Molecular Mechanisms Underlying the Progression of miR-21 in Oral Cancer. *World J. Surgical Oncology* 2025, 23 (1), 73.

27. Nguyen, H.T.; Salah, E.O.K.; Truc, L.N.; Kamrul, H.S.; Roselyn, L.M; Humaira, S.; Duy, N.D. MiR-21 in the Cancers of the Digestive System and Its Potential Role as a Diagnostic, Predictive, and Therapeutic Biomarker. *Biology* 2021, 10 (5). https://doi.org/10.3390/biology10050417.

28. Kura, B.; Barbora, K.; Yvan, D.; Monika, B. Potential Clinical Implications of miR-1 and miR-21 in Heart Disease and Cardioprotection. *Int. J. Mol. Sci.* 2020, 21(3). https://doi.org/10.3390/ijms21030700.

29. Surina, R.A.F.; Lucia, S.; Raffaele, M.; Giuseppe, P.; Michelangela, B. miR-21 in Human Cardiomyopathies. *Frontiers in Cardiovascular Medicine* 2021, 8 (October):767064.

30. Jayawardena, E.; Lejla, M.; Gregoire, R.; Mansoureh, E. Role of miRNA-1 and miRNA-21 in Acute Myocardial Ischemia-Reperfusion Injury and Their Potential as Therapeutic Strategy. *Int. J. Mol. Sci.* 2022, 23(3), 1512.

31. Sessa, F.; Monica, S.; Massimiliano, E.; Giuseppe, C.; Cristoforo, P. miRNA Dysregulation in Cardiovascular Diseases: Current Opinion and Future Perspectives. *Int. J. Mol. Sci.* 2023, 24(6), 5192.

32. Zhu, S.; Hailong, W.; Fangting, W.; Daotai, N.; Shijie, S.; Yin-Yuan, Mo. MicroRNA-21 Targets Tumor Suppressor Genes in Invasion and Metastasis. *Cell Research* 2008, 18 (3), 350–59.

33. Zhang, J.G.; Jian-Jun, W.; Feng, Z.; Quan, L.; Ke, J.; Guang-Hai, Y. MicroRNA-21 (miR-21) Represses Tumor Suppressor PTEN and Promotes Growth and Invasion in Non-Small Cell Lung Cancer (NSCLC). *Clinica Chimica Acta; Int. J. Clin. Chem.* 2010, 411 (11-12), 846–52.

34. Rhim, J.; Woosun, B.; Yoona, S.; Jong, H.K. From Molecular Mechanisms to Therapeutics: Understanding MicroRNA-21 in Cancer. *Cells* 2022, 11(18). https://doi.org/10.3390/cells11182791.

35. Liu, G.; Arnaud, F.; Yanping, Y.; Jadranka, M.; Qiang, D.; Victor, J.T.; Naftali, K.; Edward, A. miR-21 Mediates Fibrogenic Activation of Pulmonary Fibroblasts and Lung Fibrosis. *J. Exp. Med.* 2010, 207(8), 1589–97.

36. Sheedy, F.J. Turning 21: Induction of miR-21 as a Key Switch in the Inflammatory Response." *Frontiers in Immunology* 2015, 6 (January),19.

37. Lopez, M.S., Robert, J.D.; Raghu, V. The microRNA miR-21 Conditions the Brain to Protect against Ischemic and Traumatic Injuries. *Conditioning Medicine* 2017, 1(1), 35–46.

38. Bulygin, K.V.; Narasimha, M.B.; Aigul, R.S.; Vladimir, N.N.; Ilgiz, G.; Ozal, B.; Leila, R.A. Can miRNAs Be Considered as Diagnostic and Therapeutic Molecules in Ischemic Stroke Pathogenesis?—Current Status. *Int. J. Mol. Sci.* 2020, 21(18), 6728.

39. Correia, S.; Marta, N.C.; Cyril, S.; Monika, G.; Sophie, C.; Christine, M.; Dobrochna, D. Mir-21 Suppression Promotes Mouse Hepatocarcinogenesis. *Cancers* 2021, 13(19), 4983.

40. Sivamaruthi, B.S.; Neha, R.; Mehul, C.; Sankha, B.; Bhupendra, G.P.; Gehan, M.E.; Chaiyavat, C. NF-κB Pathway and Its Inhibitors: A Promising Frontier in the Management of Alzheimer's Disease. *Biomedicines* 2023, 11(9). https://doi.org/10.3390/biomedicines11092587.

41. Zhang, W.C.; Nicholas, S.; Fareesa, A.; Cerena, M.; Luis, S.; Paul, J.A.C.; John, M.A.;, Aaron, N.H,; Frank, J.S. MicroRNA-21 Guide and Passenger Strand Regulation of Adenylosuccinate Lyase-Mediated Purine Metabolism Promotes Transition to an EGFR-TKI-Tolerant Persister State. *Cancer Gene Therapy* 2022, 29 (12), 1878–94.

42. Gilyazova, I.; Dilara, A.; Evelina, K.; Ruhi, S.; Artur, M.; Elizaveta, I.; Ksenia, B. MiRNA-146a-A Key Player in Immunity and Diseases. *Int. J. Mol. Sci.* 2023, 24(16). https://doi.org/10.3390/ijms241612767.

43. Gronau, L.; Katharina, B.; Stefan, Z.; Ralf, S. Dual Role of microRNA-146a in Experimental Inflammation in Human Pulmonary Epithelial and Immune Cells and Expression in Inflammatory Lung Diseases. *Int. J. Mol. Sci.* 2024, 25(14). https://doi.org/10.3390/ijms25147686.

44. Komoll, R.M.; Qingluan, H.; Olaniyi, O.; Lena, D.; Qinggong, Y.; Yu, X.; Hsin-Chieh, T. MicroRNA-342-3p Is a Potent Tumour Suppressor in Hepatocellular Carcinoma. *J. Hepatology* 2021, 74 (1), 122–34.

45. Dakterzada, F.; Iván, D.B.; Adriano, T.; Albert, L.; Gerard, T.; Leila, R.; David, G.C. Reduced Levels of miR-342-5p in Plasma Are Associated With Worse Cognitive Evolution in Patients With Mild Alzheimer's Disease. *Front. Aging Neurosci.* 2021, 13 (August), 705989.

46. Taghehchian, N.; Yalda, S.; Amirhosein, M.; Amir S.Z.; Samaneh, B.N.; Meysam, M. Molecular Biology of microRNA-342 during Tumor Progression and Invasion. *Pathology, Research and Practice* 2023, 248 (August), 154672.

47. Cheng, S.; Yanxiang, C.; Lin, F.; Xiaofeng, M.; Yuzhong, H. T2DM Inhibition of Endothelial miR-342-3p Facilitates Angiogenic Dysfunction via Repression of FGF11 Signaling. *Biochemical and Biophysical Research Communications* 2018, 503(1), 71–78.

48. Roy, B.; Erica, L.; Teresa, L.; Maria, R. Role of miRNAs in Neurodegeneration: From Disease Cause to Tools of Biomarker Discovery and Therapeutics. *Genes* 2022, 13(3), 425.

49. Ma, Y.M.; Lan, Z. Mechanism and Therapeutic Prospect of miRNAs in Neurodegenerative Diseases." *Behavioural Neurology* 2023, (November), 8537296.